\documentclass[11pt]{article}
\usepackage{times}
\usepackage{graphics}
\usepackage{color}
\usepackage{amssymb}
\usepackage{amsmath}
\usepackage{ifthen}
\usepackage{calc}
\usepackage{float}
\usepackage{hyperref}
\usepackage{color}

\newtheorem{theorem}{Theorem}

\newlength{\Ainlength}
\newlength{\Ainindent}
\newlength{\Aintemp}

\newcommand{\Ainnum}[2]{\makebox[\Ainindent][r]{#2}\hspace*{0.1in}\setlength{\Aintemp}{0pt}\addtolength{\Aintemp}{#1\Ainlength} \hspace*{\Aintemp}}
\renewcommand{\vec}[1]{\ensuremath{\mathbf{#1}}}
\newcommand{\data}{\ensuremath{(\vec{y},\vec{w})}}
\newcommand{\mean}[1]{\ensuremath{\mathsf{mean}\ifthenelse{\equal{#1}{}}{}{(#1)}}}
\newcommand{\meanerr}[1]{\ensuremath{\mathsf{mean\_err}\ifthenelse{\equal{#1}{}}{}{(#1)}}}

\newcommand{\elow}{\ensuremath{\epsilon_{\mathsf{low}}}}
\newcommand{\ehigh}{\ensuremath{\epsilon_{\mathsf{high}}}}
\newcommand{\eopt}{\ensuremath{\epsilon_{\mathsf{opt}}}}
\newcommand{\denv}[1]{\ensuremath{\mathsf{denv}\ifthenelse{\equal{#1}{}}{}{(#1)}}}
\newcommand{\uenv}[1]{\ensuremath{\mathsf{uenv}\ifthenelse{\equal{#1}{}}{}{(#1)}}}
\newcommand{\myweb}{www.eecs.umich.edu/{\,\raisebox{-0.5ex}{\textasciitilde}qstout/}abs/}
\definecolor{linkcolor}{rgb}{0.0,0.0,0.5}

\floatstyle{plain}
\newfloat{algorithm}{tbp}{algorithm}

\floatname{algorithm}{Algorithm}

\begin{document}

\begin{center}
\textbf{\Large $\mathbf{L_\infty}$ Isotonic Regression for Linear, Multidimensional, and Tree Orders}
\bigskip
\bigskip

{\large Quentin F. Stout}
\medskip

qstout@umich.edu
\smallskip

University of Michigan\\
Ann Arbor, MI
\end{center}

\medskip

\subsubsection*{Abstract}
Algorithms are given for determining $L_\infty$ isotonic regression of weighted data. 
For a linear order, grid in multidimensional space, or tree, of $n$ vertices, optimal algorithms are given, taking $\Theta(n)$ time.
These improve upon previous algorithms by a factor of $\Omega(\log n)$.
For vertices at arbitrary positions in $d$-dimensional space a $\Theta(n \log^{d-1} n)$ algorithm employs iterative sorting to yield the functionality of a multidimensional structure while using only $\Theta(n)$ space.
The algorithms utilize a new non-constructive feasibility test on a rendezvous graph, with bounded error envelopes at each vertex.

\medskip

\noindent
\textbf{Keywords}: weighted $L_\infty$ isotonic regression, shape-constrained nonparametric regression, linear order, multidimensional domination, tree, rendezvous graph,  coordinate-wise ordering

\section{Introduction}    \label{sec:intro}

This paper gives efficient algorithms for determining optimal $L_\infty$ isotonic regression functions for weighted data.
For example, consider predicting weight as a function of height and S $<$ M $<$ L $<$ XL shirt size.
Average weight is an increasing function of height and of shirt size, but there may be no assumptions about the relative weights of people shorter and with a larger shirt size vs.\ taller and smaller shirt size.
A parametric model, such as linear regression, may not be justified, and would require a metric on shirt sizes, not just an ordering.
Isotonic regression is useful here since it merely assumes a direction on each variable.

Isotonic regression is a fundamental nonparametric regression, long used for numerous applications~\cite{BarlowetalBook,RobertsonWrightDykstra} and mathematically equivalent problems~\cite{KaufmanTamir}, and is getting increased attention in machine learning and data mining due to its flexibility and minimal assumptions~\cite{CaruanaNicul06,Chakrabartietal07,Gamarnik,Isotron,KotlowskiSlowinskiClassification,IntervalRank,PuneraGhosh08,RadeetalRelabel,VelikovaDaniels}.
For example, the Isotron algorithm provably learns single index models~\cite{Isotron}.
As another example, a classification system may have some confidence that an item in an image is squirrel, of it being a rat, a mammal, etc., with the isotonic requirement that as one moves up the taxonomy tree the confidence does not decrease~\cite{deKampetalClassificationTree,PuneraGhosh08}.

\begin{figure}
\begin{center}
\resizebox{!}{1.3in}{\includegraphics{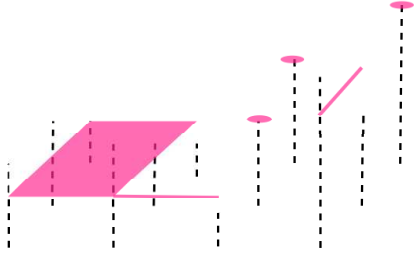}}
\end{center}

\vspace*{0.1in} 
\hrulefill
\vspace*{-0.45in}

\caption{Isotonic Regression on 4$\times$3 Grid} \label{fig:Isotonic2D}
\end{figure}

Formally, given a directed acyclic graph (dag) $G=(V,E)$, for vertices $u,v \in V$,
$u \prec v$ iff there is a path in $G$ from $u$ to $v$.
A real-valued function $f$ on $V$ is \textit{isotonic} iff whenever $u \prec v$ then $f(u) \leq f(v)$.
By \textit{weighted data} \data\ on $G$ we mean functions $y, w$ on $V$ where $y(v)$ is an arbitrary real value and $w(v)$ (the weight) is nonnegative, for $v \in V$.
Given weighted data \data,\ an isotonic function $f$ on $V$ is an \textit{$L_p$ isotonic regression} iff it minimizes the weighted $L_p$ error
$$
\begin{array}{ll}
  \left(\sum_{u \in V}\, w(u) |y(u) -f(u)|^p\right)^{1/p} & 1 \leq p < \infty \medskip \\
  \max_{u \in V} w(u)\, |y(u) - f(u)|                     & p = \infty
\end{array}
$$
among all isotonic functions.
See Figure~\ref{fig:Isotonic2D}.
Note that when $V$ is points on the real line, $f$ is not necessarily defined at points not in $V$.
For example, if all weights are the same and the $(x,y)$ values are (1,7), (2,5), (3,8),
then for all $p$ an optimal regression is 6 on [1,2] and 8 at 3, but for any $x \in (2,3)$ and $y \in [6,8]$ there is an optimal isotonic regression $f$ for which $f(x)=y$.
Further, while isotonic regression is unique when $1 < p < \infty$, for $p=\infty$ it is not necessarily unique. 
For example, given values 4, 1, 3, with weights 1, 1, 1, on 0, 1, 2 with the natural ordering, a function is an $L_\infty$ isotonic regression iff its values are 2.5, 2.5, $c$, for $c \in [2.5, 4.5]$.

For a dag of $n$ vertices and $m$ edges, previous algorithms for weighted $L_\infty$ regression were $\Omega(m \log n)$ no matter what the dag was~\cite{KaufmanTamir,LinetalInverseSort,LiuUbhaya,QLinfty}, though a randomized algorithm taking $\Theta(m)$ expected time appears in~\cite{YaleIsoReg}.
Here we give $\Theta(n)$ algorithms for the dags of most interest, namely linear, tree, and multidimensional grids.
We also give an algorithm for arbitrary points in $d$-dimensional space with component-wise ordering, $d \geq 2$, taking
$\Theta(n \log^{d-1} n)$ time and only $\Theta(n)$ space.
For all $1 \leq p \leq \infty$, previous algorithms for $L_p$ isotonic regression for such inputs required $\Omega(n \log^{d-1} d)$ space and more time.

\section{Preliminaries}   \label{sec:prelim}

For linear orders it is well known that the $L_1$, $L_2$, and $L_\infty$ isotonic regressions can easily be found in $\Theta(n \log n)$, $\Theta(n)$ time, and $\Theta(n \log n)$ time, respectively, by using ``pool adjacent violators'' (PAV) where whenever adjacent level sets are not isotonic they are pooled together into a larger set where all the regression values are the same.
Algorithms for multidimensional orderings have concentrated on 2 dimensional grids~\cite{BeranDumbgen,DykstraHewettRobertson,DykstraRobertson,Gebhardt,MeyerInference,QianEddy,RobertsonWrightMultiple,SpougeWanWilber}, with the fastest taking $\Theta(n \log n)$ time for $L_1$~\cite{QPartition},
$\Theta(n^2)$ for $L_2$~\cite{SpougeWanWilber}, and $\Theta(n \log n)$ for $L_\infty$~\cite{QLinfty}.
For grids of dimension $> 2$, and $d$-dimensional points in arbitrary positions, $d \geq 2$, the fastest algorithms just apply the fastest algorithm for arbitrary dags, which takes $\Theta(nm + n^2 \log n)$ time for $L_1$~\cite{AHKW} ($\Theta(\min\{nm+n^2 \log n,\, n^{2.5} \log n\})$ if the data is unweighted~\cite{QRendez}), $\Theta(nm \log n)$ for $L_2$~\cite{HochQuey}, and $\Theta(m \log n)$ for $L_\infty$~\cite{QLinfty} ($\Theta(m)$ if the data is unweighted).

Given data \data\ on dag $G=(V,E)$, for $u, v \in V$, with $u \prec v$ and $y(u) \geq y(v)$, let $\mean{u,v}= (y(u)w(u) + y(v)w(v))/(w(u)+w(v))$
and $\meanerr{u,v} = w(u)w(v)(y(u)+y(v))/(w(u)+w(v))$.
For any isotonic function $f$, the weighted error of at least one of $f(u)$ and $f(v)$ is $\geq \meanerr{u,v}$, which is minimized when $f(u) = f(u) = \mean{u,u}$.
Because of the isotonic restriction, a larger value at $u$ forces a larger value at $u$, increasing the error there, and, similarly, a smaller value at $v$ increases the error at $u$.
The most widely known $L_\infty$ isotonic regression $f$ is given by 
$f(x) = \mean{u^\prime,v^\prime}$, where

\vspace{0.05in}
\centerline{$(u^\prime,v^\prime) = \mathrm{argmax}\{\meanerr{u,v}: u, v \in V,\, u \preceq x \preceq v,\, y(u) \geq y(v)\}$}
\vspace{0.05in}
\noindent Note that the optimal regression error, \eopt, is $\max\{\meanerr{u,v}: u \preceq v,\, y(u) \geq y(v)\}$.

A simplistic use of this could take $\Theta(n^3)$ time, so more efficient methods are used instead.
Given an $\epsilon > 0$, to decide if $\epsilon \geq \eopt$, for $x \in V$ let $h(x) = y(x)-\epsilon/w(x)$, i.e., the smallest value at $x$ with weighted error no more than $\epsilon$, and let $g(u) = \max\{h(x): x \preceq u\}$.
$g(u)$ is the smallest possible value at $u$ of any isotonic function with error $\leq \epsilon$ on $u$ and its predecessors, and
thus there is an isotonic function with error $\leq \epsilon$ iff $g$ is one, i.e., iff $g(u) \leq y(u)+\epsilon/w(u)$ for all $u \in V$.
This is usually called a feasibility test, but here will be called a \textit{feasibility construction}.
It can be computed in $\Theta(m)$ time via topological sort.

For the $L_\infty$ metric, isotonic regression and related nonparametric regressions such as $b$-step (where the regression is piecewise constant with at most $b$ pieces), researchers have used an approach based on searching through \meanerr{} values to find \eopt, using a feasibility construction to guide the search and to generate the final regression~\cite{DiazBanezLinearDecide,FournierVigneronLinftyParametric,GuhaShimLinftyHistogram,KarrasetalHistogramDuality,KaufmanTamir,LiuRandomizedLinftyReduced,QLinfty}.
$\Omega(\log n)$ feasibility constructions are performed, so the time is $\Omega(m \log n)$.

This paper gives a technique for determining \eopt\ more rapidly, using a nonconstructive feasibility test (``pairwise feasibility test'', Sec.~\ref{sec:meanerror}).
It utilizes a ``rendezvous graph'' (Sec.~\ref{sec:rendezvous}) as a succinct representation of the transitive closure of the original dag.
This graph has $\Theta(\log n)$ levels, with the base corresponding to the vertices of the original dag.
The algorithm traverses the rendezvous graph level by level, with
``bounded error envelopes'' (Sec.~\ref{sec:bounded}) used to generate \meanerr{} values and for the pairwise error test.

\subsection{Bounded Error Envelopes} \label{sec:bounded}

\begin{figure}
\begin{center}
\resizebox{!}{1.3in}{\includegraphics{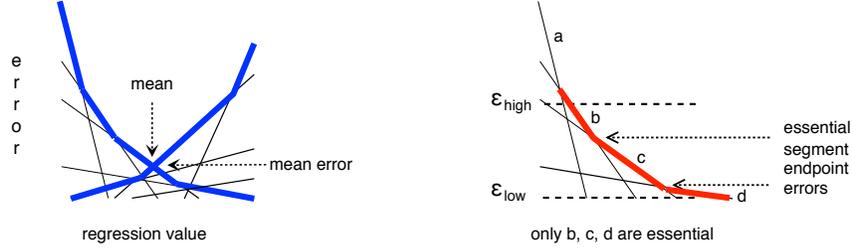}}

\end{center}

\vspace*{0.1in} 
\hrulefill
\vspace*{-0.45in}

\caption{Standard error envelopes and a bounded downward error envelope} \label{fig:envelope}
\end{figure}

There is a geometric interpretation of \mean{} and \meanerr{}:
for a weighted value $(a,b)$, the ray in the upper half plane that starts at $(a,0)$ with slope $b$ gives the error of using a regression value greater than $a$ (this will be called the \textit{upward ray}) and the ray that starts at $(a,0)$ with slope $-b$ gives the error of using a smaller value (this will be called the \textit{downward ray}).
If $y_1 < y_2$ then \mean{(y_1,\!w_1),(y_2,\!w_2)} and \meanerr{(y_1,\!w_1),(y_2,\!w_2)} are given by the intersection of the upward ray of $(y_1,w_1)$ and the downward ray of $(y_2,w_2)$.
Given disjoint sets $V_1, V_2 \subseteq V$, let $s=\max\{\meanerr{v_1,v_2}: v_1 \in V_1, v_2\in V_2, y(v_1) \geq y(v_2)\}$.
If every element of $V_1$ precedes every element of $V_2$, then any isotonic function must have error at least $s$.
To determine $s$ let $D$ be the set of downward rays corresponding to the elements of $V_1$ and $U$ the set of upward rays corresponding to elements of $V_2$.
The \textit{downward envelope} of $V_1$ is the piecewise linear function which is the pointwise maximum over all rays in $D$, and the \textit{upward envelope} of $V_2$ is the pointwise maximum of all rays in $U$
(see Figure~\ref{fig:envelope}).
The value of $s$ is the error of the intersection of the downward envelop of $V_1$ and the upward envelope of $V_2$ (if they don't intersect then $s=0$), and the ordinate of this intersection is the regression value achieving $s$.

While error envelopes have been used by many authors, in general not all of the segments in the envelopes are necessary.
Suppose bounds \elow, \ehigh\ are known, where $\elow \leq \eopt < \ehigh$.
Then only the portions of the envelopes representing errors in $(\elow,\ehigh)$ are needed to determine \eopt.
The segments of the envelope representing errors in $(\elow,\ehigh)$ are \textit{essential}, and the others are inessential and can be pruned.
The essential segments form a \textit{bounded envelope}, as shown in Figure~\ref{fig:envelope}.
The interval $[\elow,\ehigh]$ is continually narrowed.
However, this is only used to reduce the number of essential segments at the current level or above:
nodes at lower levels are never revisited.

Bounded envelopes will be associated with vertices of the rendezvous graph described below, where the downward bounded envelope at vertex $u$ is denoted \denv{u} and its upward bounded envelope is \uenv{u}.
Since we will always have an upper bound $\ehigh > \eopt$, and the intersection of \denv{u} and \uenv{u} is no greater than \eopt, with bounded envelopes we can determine the intersection of \denv{u} and \uenv{u} if it is $\geq \elow$, and if it is $< \elow$ then we can determine that fact though not necessarily its (not needed) value.
Bounded envelopes are stored as simple lists ordered by slope, and all operations on bounded envelopes
can be done in time linear in the number of segments.

\subsection{Rendezvous Graphs}  \label{sec:rendezvous}

\begin{figure}
\hspace{0.25in}
\begin{minipage}[t]{2.55in}
\resizebox{1.6in}{!}{\includegraphics{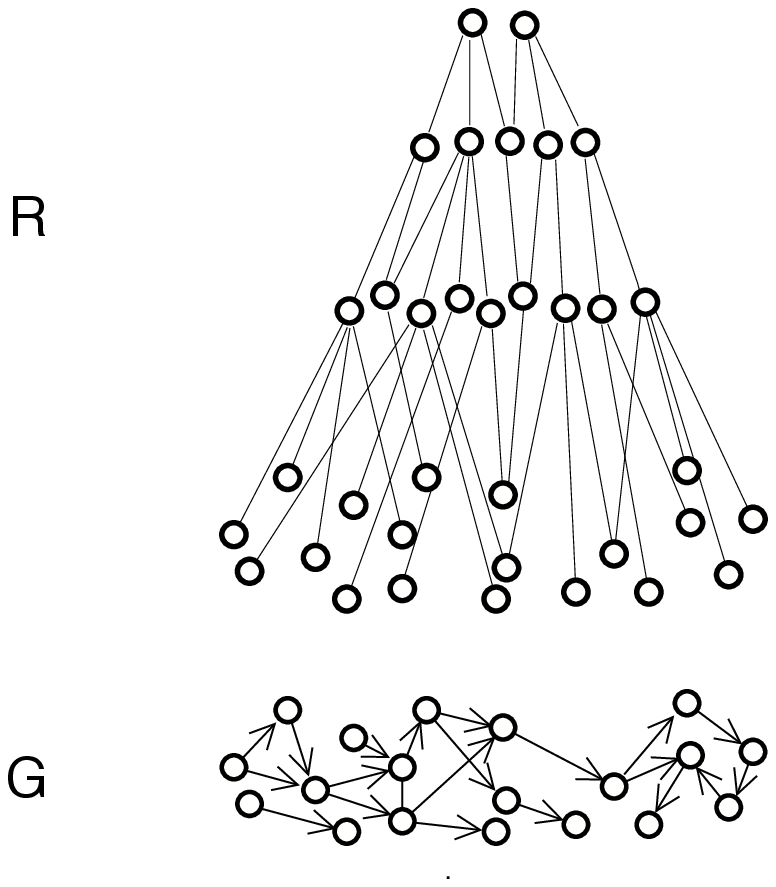}}
\end{minipage}
\begin{minipage}[t]{3.55in}
\resizebox{3.5in}{!}{\includegraphics{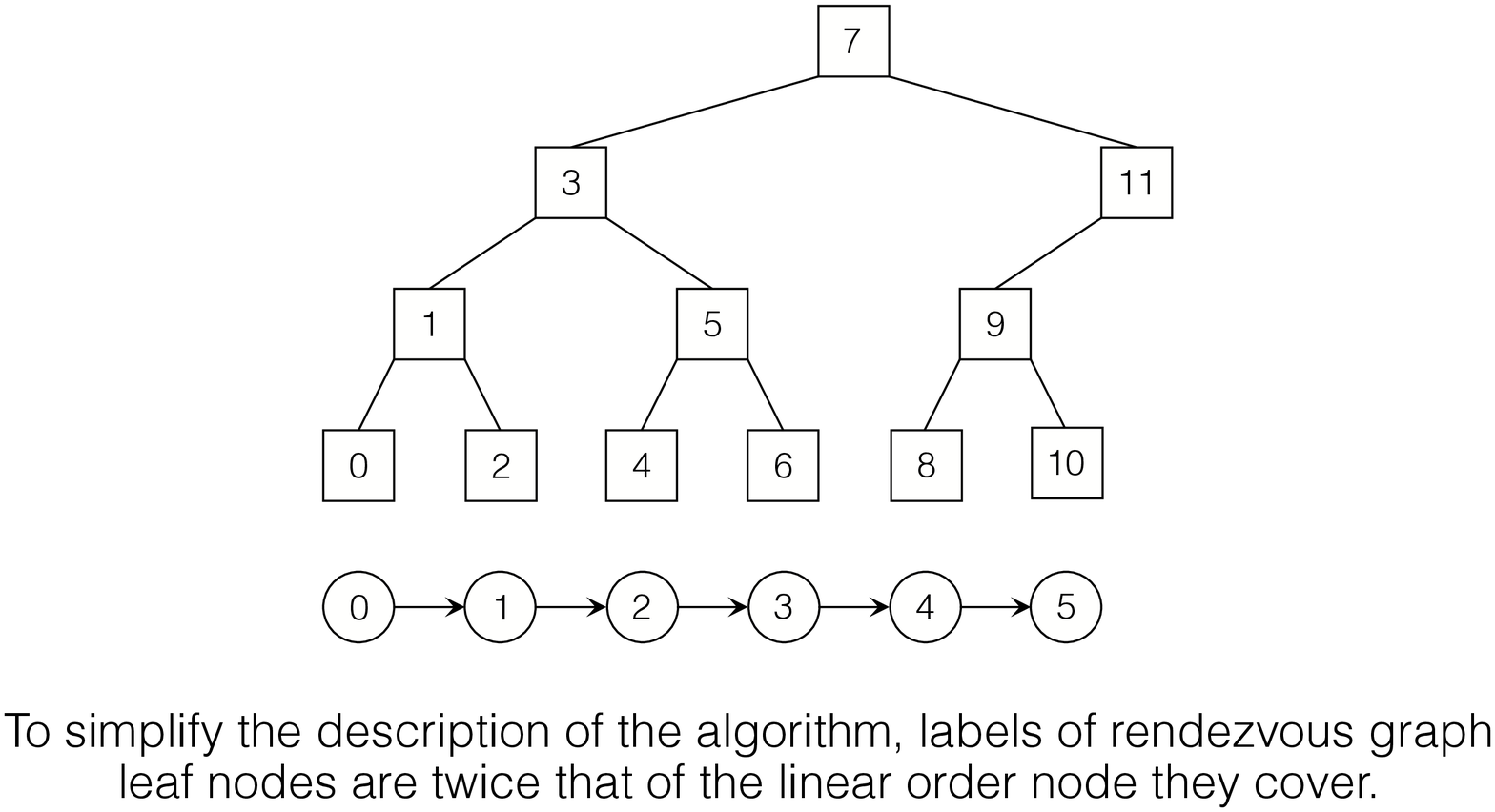}}
\end{minipage}

\vspace*{0.3in} 
\hrulefill
\vspace*{-0.4in}

\caption{Rendezvous graphs: generic, and for linear order, $n$=6} \label{fig:linear}
\end{figure}

For a dag $G=(V,E)$, a \textit{rendezvous graph} $R=(V_R,E_R)$ is used to help speed up the calculations.
These were introduced in~\cite{QRendez}, though here they are modified slightly.
The nodes of $R$ have height $0, \ldots, h$, for some $h \geq 1$, where the leaves are at height 0 and correspond to the vertices of $G$.
A node at height $i$ only has edges to nodes at height $i-1$ and $i+1$.
For dags more complicated than a linear order the rendezvous graph is not a tree, but we will still call the adjacent nodes below children and adjacent nodes above parents (see Figure~\ref{fig:linear}).
A node $r \in V_R$ \textit{covers} a set $S(r) \subset V$, where $S(r)$ corresponds to the set of leaf nodes in the descendants of $r$.
A node $r$ may have several children, but there is a unique small child, denoted $r_S$, and a unique large child, denoted $r_L$, such if $v,u \in V$ and $r_S$ covers $u$ and $r_L$ covers $v$, then $u \prec v$ in $G$.
Further, for every pair of vertices $u,v \in V$, if $u \prec v$ then there is a vertex $r \in V_R$ such that $u$ is covered by $r_S$ and $v$ is covered by $r_L$.
This is their \textit{rendezvous node}.
Here their rendezvous node is unique, but the algorithm only depends upon the existence of at least one rendezvous for them.

To explain the presence of other children, suppose $r$ covers a rectangular matrix, with children nodes $r_{00}$, $r_{01}$, $r_{10}$, and $r_{11}$, each covering a quadrant of the rectangle.
If $r_{00}$ covers the lower left quadrant (in standard ordering), then it is the small child, and if $r_{11}$ covers the upper left quadrant it is the large child.
Both of $r_{01}$ and $r_{10}$ cover vertices that are successors of some vertices covered by  $r_{00}$ but not all, and vertices that are predecessors of some elements covered by $r_{11}$ but not all.
Vertices in $V$ covered by $r_{00}$ or $r_{11}$ rendezvous with those covered by $r_{01}$ or $r_{10}$ at lower levels.
However, data values in all 4 quadrants are needed to construct $r$'s bounded error envelopes to be used at higher levels.

For every vertex $r \in V_R$, the intersection of \denv{r_S} and \uenv{r_L} gives
\vspace*{-0.1in}
$$
\epsilon_r = \max\{\meanerr{v,w}: v \in V \mathrm{~covered~ by~} r_S,~ w \in V \mathrm{~covered~ by~} r_L,~ y(v) \geq y(w)\}
\vspace*{-0.1in}
$$
\meanerr{} values involving vertices not in $r_S$ nor $r_L$, are ignored.
Since every pair $u \prec v$ in $G$ rendezvous somewhere, $\eopt = \max\{\epsilon_r: r \in V_R \}$.
Whenever $\epsilon_r$ is calculated we set $\elow=\max\{\elow, \epsilon_r\}$ and never revisit $r$.

\subsection{Pairwise Feasibility Test and Reducing the Number of Segments} \label{sec:meanerror}

A \textit{pairwise feasibility test} is given $\delta > \elow$, and determines if $\meanerr{u,v} \leq \delta$ for all pairs $u \prec v$ where $y(u) \geq y(v)$.
If true, $\ehigh = \min\{\ehigh, \delta\}$.
It can be proven by showing $\epsilon_r \leq \delta$ for all rendezvous nodes $r$.
Let $D(r,x) = \denv{r}^{-1}(x)$, i.e., the regression value of the point on \denv{r} with error $x$, and $U(r,x) = \uenv{r}^{-1}(x)$ be the corresponding value for the upward envelope.
Then $\epsilon_r \leq x$ iff $U(r_S,x) \leq D(r_L,x)$.
A test is initiated at level $h$ after $\epsilon_r$ has been determined for all vertices $r$ at level $h$ and below, in which case \elow\ is the largest such value
($\epsilon_r$ isn't computed exactly if $U(r,\elow) \leq D(r,\elow)$ since $\epsilon_r \leq \elow$).
Thus we only need to determine if $\epsilon_r \leq \delta$ for vertices at level $h+1$ and above.
First, for all vertices $v \in V_R$ at height $h$, construct \denv{v} and \uenv{v} from the union of the envelopes of all of $v$'s children.

For a pairwise feasibility test of $\elow < \delta < \ehigh$ initiated at level $h$:
\begin{enumerate}
\item \vspace*{-0.05in} For all vertices $v$ at height h, compute $D(v,\delta)$ and $U(v,\delta)$.
\item \vspace*{-0.08in}Level by level, starting at level $h+1$, for a vertex $r$, if $D(r_S,\delta) > U(r_L,\delta)$ then the test fails.
Otherwise, $D(r,\delta) = \max\{D(v,\delta): v \mathrm{~a~child~of~} r\}$ and $U(r,\delta) = \min\{U(v,\delta): v ~\mathrm{~a~child~of~} r\}$.
\end{enumerate}
The pairwise feasibility test succeeds iff it does not fail anywhere, in which case $\ehigh = \delta$.
Its time is the time to compute the bounded envelopes at level $h$, which is linear in the number of envelope segments at that level, plus time linear in the sum, over all nodes at height $h+1$ and above, of the number of children.
The algorithms insure that both terms are $O(n/2^h)$.

To reduce the time the number of envelope segments is continually reduced.
After \denv{r} and \uenv{r} are created for all nodes $r$ at height $h$, for a bounded upward or downward envelope at a node, consider the errors of the endpoints of its essential segments (see Fig.~\ref{fig:envelope}) and let $E$ be the union of these errors over all nodes at level $h$ ($E$ is a multiset).
Then either $\eopt \in E$ or there are two consecutive errors $\epsilon_1$, $\epsilon_2 \in E$ such that $\epsilon_1 < \eopt < \epsilon_2$ (or \eopt\ is less than the smallest, or greater than the largest, value in $E$).
By taking the median error $\delta$ in $E$ and using a pairwise feasibility test to determine if $\delta \geq \eopt$ one can eliminate 1/2 of the endpoint errors in $E$ and eliminate corresponding segments in the envelopes.
This is not quite the same as eliminating 1/2 the segments since each bounded envelope has at least one segment, hence the number of essential segments is $2x + |E|$, where $x$ is the number of vertices at level $h$.
\medskip

\noindent \textbf{Observation 1:}
Let $c < 1$ be given.
Suppose there are $\leq 3x$ essential segments at level $h$, each rendezvous node has at most $k$ parents (and hence each bounded envelope from level $h$ is used at most $k$ times to create envelopes at level $h+1$),
and the number of nodes at level $h+1$ is $\leq cx$.
Then initially there are at most $3kx$ essential segments at level $h+1$, and hence at most this many segment endpoint errors.
Using the median $\lceil \lg 3k/c \rceil$ times to reduce the number of segments errors results in at most $cx$ segment endpoint errors, and thus at most $3cx$ essential segments, at level $h+1$.
Thus if we always use the median this many times at every level and start with $2n$ segments at the base there will be $ \leq c^\ell n$ essential segments at level $\ell$.

\section{$d$-Dimensional Points in Grids and Arbitrary Position}  \label{sec:grids}

For points in $d$-dimensional space, $d \geq 1$, the component-wise ordering, also known as \textit{domination} or \textit{product order}, is used, i.e., $(a_1,\ldots,a_d) \prec (b_1,\ldots,b_d)$ iff $a_j \leq b_j$ for all $1 \leq j \leq d$.
Since it is only the ordering of the independent variable that is important, not their values, for linear orders we assume the values are $\{0,\ldots,n-1\}$.
A \textit{$d$-dimensional grid} of size $n_1\times \ldots \times n_d$\, has vertices $(i_1, \ldots i_d)$, where $0 \leq i_j \leq n_j-1$ for all $1 \leq j \leq d$.
To avoid degeneracy we assume that $n_j \geq 2$ for all $1 \leq j \leq d$.
For points in arbitrary position the $i^\mathrm{th}$ coordinate, $1 \leq i \leq d$, has values $0\ldots n_i-1$, where each value appears at least once.
If the original coordinates are different they can be converted to this form in $\Theta(n \log n)$ time, but this is only needed for points in arbitrary position.

For grids the input is a $n_1 \times \ldots \times n_d$ array, but for points in arbitrary position
we merely assume the data is given in a list or linear array format
since it may be that $n \ll n_1 \times \ldots \times n_d$.

\begin{theorem}   \label{thm:points}
Given weighted data \data\ on a set of\, $n$ vertices in $d$-dimensional space with component-wise ordering, $d \geq 1$, an $L_\infty$ isotonic regression of the data can be determined in
\begin{enumerate}
\item[a)] \vspace*{-0.07in} $\Theta(n)$ time and space if the vertices form a grid, and
\item[b)] \vspace*{-0.07in} $\Theta(n \log^{d-1} n)$ time and $\Theta(n)$ space if the vertices are in arbitrary positions, \label{trial}
\end{enumerate}
where the implied constants depend upon $d$.
\end{theorem}

\vspace*{-0.07in}\noindent
a) is proven in Sec.~\ref{sec:linear} for linear orders and in Sec.~\ref{sec:multi} for $d$-dimensional grids.
b) is proven in Sec.~\ref{sec:arbitrary}.

\subsection{Linear Orders}  \label{sec:linear}

Let $G=(V,E)$ be a linear order with vertices $\{0, \ldots, n-1\}$.
The rendezvous graph $R=(V_R,E_R)$ for $G$ is a simple binary tree (see Figure~\ref{fig:linear}).
The only unusual aspect is that the vertices in $V_R$ at height 0 have labels twice that of the ones they correspond to in $V$, which is used merely to simplify the description of the algorithm.
A vertex $i$ in $V_R$ at height $h \geq 1$ has two children $i \pm 2^{h-1}$, though the larger child will be absent if $i > 2n-2$.
The maximum height is $L=\lceil \lg n \rceil$ and $|V_R| < 2^{L +1} + L < 3n$.

\begin{algorithm}
\setlength{\Ainindent}{0.45in}

\Ainnum{0}{1}$\mathsf{ for~ every~ rendezvous~ vertex~ r~ at~ height~ 0~ initialize~ \denv{r},~ \uenv{r}~~ \{single~ rays\}}$\\
\Ainnum{0}{2}$\mathsf{ \elow=0;~~ \ehigh=\infty}$\\
\Ainnum{0}{3}$\mathsf{ for~ h~ =1 ~to~ \lceil \lg n \rceil~~~ \{number~ of~ segments~ in~ envelopes~ at~ level~ h-1~ is~ \leq \lceil 3n/2^{h-1} \rceil\}}$\\
\Ainnum{1}{4}$\mathsf{   E = \emptyset~~ \{E~ is~ multiset~ of~ segment~ endpoint~ errors\}}$\\
\Ainnum{1}{5}$\mathsf{   for~ every~ vertex~ r~ at~ height~ h~~~ \{modifications~ needed~ for~ r > 2n-2~ since~ no~ large~ child\}}$\\
\Ainnum{2}{6}$\mathsf{   r_S = r-2^{h-1};~~ r_L = r+2^{h-1}~~ \{small~ and~ large~ child,~ respectively\} }$\\
\Ainnum{2}{7}$\mathsf{      \elow = \max\{\elow,~ error~ of~ intersection~ of~ \denv{r_S} ~and~ \uenv{r_L}} \} $\\
\Ainnum{2}{8}$\mathsf{      \denv{r} = merge(\denv{r_S},\denv{r_L});~~ \uenv{r} = merge(\uenv{r_S},\uenv{r_L}) }$\\
\Ainnum{2}{9}$\mathsf{      add~ endpoint~ errors~ of~ \denv{r}~ and~ \uenv{r}~ that~ are~ in~ (\elow,\ehigh)~ to~ E}$\\
\Ainnum{1}{10}$\mathsf{     repeat~ 3~ times~~~ \{insures~ final~ |E| \leq \lceil n/2^h \rceil\}}$\\
\Ainnum{2}{11}$\mathsf{       \delta~ =~ median~ error~ of~ values~ in~ E}$\\
\Ainnum{2}{12}$\mathsf{       if~ pairwise\_feasibility\_test(\delta)~ then~ \ehigh=\delta~ else~ \elow = \delta}$\\
\Ainnum{2}{13}$\mathsf{       remove~ errors~ in~ E~ outside~ (\elow,\ehigh)}$\\
\Ainnum{1}{14}$\mathsf{   for~ every~ vertex~ r~ at~ level~ h}$\\
\Ainnum{2}{15}$\mathsf{     remove~ inessential~ segments~ from~ r's~ envelopes}$\\
\Ainnum{0}{16}$\mathsf{ construct~ isotonic~ regression~ using~ \elow~~ \{at~ this~ point~ \elow = \eopt\}}$ 

\vspace*{0.3in} 
\hrulefill
\vspace*{-0.4in}

\caption{$L_\infty$ Isotonic regression of linear order using rendezvous graph} \label{alg:linear}

\end{algorithm}

Algorithm~\ref{alg:linear} gives the algorithm for isotonic regression on a linear order.
It finds the optimal regression error by using \denv{} and \uenv{} to compute the maximum error at the rendezvous nodes.
Technically line 7 is skipped if it is determined that the intersection is less than $\elow$.
For the repeat loop, lines 10--13, Observation 1 shows that only 3 iterations are required to insure that the number of essential segments at level $h$ is $\leq \lceil 3n/2^h\rceil$
(a more careful analysis shows that 2 suffice).

To determine the time of the pairwise feasibility test, in Section~\ref{sec:meanerror} it was shown that a test started at level $h$ is linear in the time for determining $\denv{r}^{-1}(\delta)$ and $\uenv{r}^{-1}(\delta)$ for all nodes at level $h$, which is linear in the number of segments in envelopes at level $h$, plus time linear in the number of nodes at higher levels.
Both terms are $\Theta(n/2^h)$, and 
thus for each iteration of the for-loop at lines 3---15 the time is $\Theta(n/2^h)$.
Summing over all levels gives $\Theta(n)$, proving Theorem~\ref{thm:points} a) for linear orders.

\subsection{Multidimensional Grids}   \label{sec:multi}

Let $G=(V,E)$ be a $n_1\times \ldots \times n_d$ $d$-dimensional grid.
Its rendezvous graph $R$ has vertices $R(n_1) \times \ldots \times R(n_d)$,
where $R(i)$ is the rendezvous graph for a linear order on $0 \ldots i-1$.
For vertex with label $s$ in $R(i)$ let $P(s)$ denote the label of its parent and $\mathrm{height}(s)$ its height
(these are independent of $i$).
For vertex $r = (r_1,\ldots,r_d)$ of $R$, $\mathrm{height}(r)\! =\! \max\{\mathrm{height}(r_j) : 1\! \leq\! j \!\leq\! d\}$.
As before, $V$ corresponds to the vertices of $R$ of height 0.

The parents of rendezvous node $x=(x_1,\ldots,x_d)$ are of the form $w=(w_1,\ldots,w_d)$, $x \neq w$, where
$$
w_j = \left\{ \begin{array}{ll}
              P(x_j) & \mbox{if height($x_j$) $> 0$}\\
              x_j \mbox{~or~} P(x_j) & \mbox{othewise}
              \end{array}
       \right.
$$
Thus $x$ has at most $2^d-1$ parents.
If any of $x$'s components are already at the maximum height in their dimension then $x$ has no parents.

For example, suppose $d=2$ and $x = (3, 8)$.
Then $\mathrm{height}(3) = 1$, $\mathrm{height}(8)=0$ (see Fig.~\ref{fig:linear}).
$x$'s parents are (7,8) and (7,9).
The ancestors of $x$ that keep 8 as their second coordinate are part of the 1-dimensional tree that covers a single row.
Meanwhile, $(7,9)$ covers a $4 \times 2$ array, its unique parent (15,11) covers a $8 \times 4$ array, etc.,
with all of its ancestors covering a rectangle with twice as many rows as columns.

Let $x=(x_1,\ldots,x_d)$, $y = (y_1,\ldots,y_d)$, $x \neq y$, be vertices in $G$.
If $x \prec y$ then their rendezvous node in $R$ is
$r=(r_1,\ldots,r_d)$ where $r_i$ is the rendezvous node of $x_i$ and $y_i$ in the linear ordering on dimension $i$.
Since $x_i \leq r_i \leq y_i$ for all $i$, $x_i$ is in the small child of $r$ and $y$ is in its large child.

A simple weak upper bound on the number of rendezvous nodes at height $h$ is if one of the coordinates is at height $h$, and all others are unconstrained.
A linear order of size $n_j$ has $ \lceil n_j/2^h\rceil < n_j/2^{h-1}$ rendezvous nodes at height $h \leq \lceil \lg n_j \rceil$ and $< 3n_j$ total rendezvous nodes, so an overestimate is
$$
\sum_{i=1}^d \frac{n_i}{2^{h-1}} \prod_{j =1,~ j\neq i}^d 3n_j = \frac{d 3^{d-1}}{2^{h-1}} \cdot n
$$
For fixed $d$ this is $ c_d\cdot n/2^h$ for a constant $c_d$.
Further, each node as at most $2^d-1$ parents, and thus Observation 1 shows that for a fixed $d$ only a constant number of pairwise feasibility tests using the median (lines 10-13 in Algorithm~\ref{alg:linear}) are needed.
Thus the total time is as claimed in Theorem~\ref{thm:points} a).

\subsection{Points in Arbitrary Position}  \label{sec:arbitrary}

For points at arbitrary positions in $d$-dimensional space their ordering is given implicitly via their coordinates, but representing this with an explicit set of edges can result in a large dag.
For example, for 2-dimensions, let $A = \{(-n,0),\, (-n+1, -1),\, (-n+2,-2), \ldots, (0, -n)\}$
and $B = \{(0,n),\, (1, n-1), \ldots (n,0)\}$.
Then every point in $A$ is dominated by every point in $B$ and representing this explicitly in a dag requires $\Theta(n^2)$ edges.
However, in a dag $G$ with vertices $A \cup B \cup (0,0)$ the ordering can be represented by an edge from each vertex of $A$ to $(0,0)$ and an edge from $(0,0)$ to each vertex of $B$, using only $\Theta(n)$ edges.
This is an order-preserving embedding of $A \cup B$ with coordinate-wise ordering into a dag
with slightly more vertices but significantly fewer edges.
$(0,0)$ is sometimes called a Steiner point.

\begin{figure}

\centerline{\resizebox{3.9in}{!}{\includegraphics{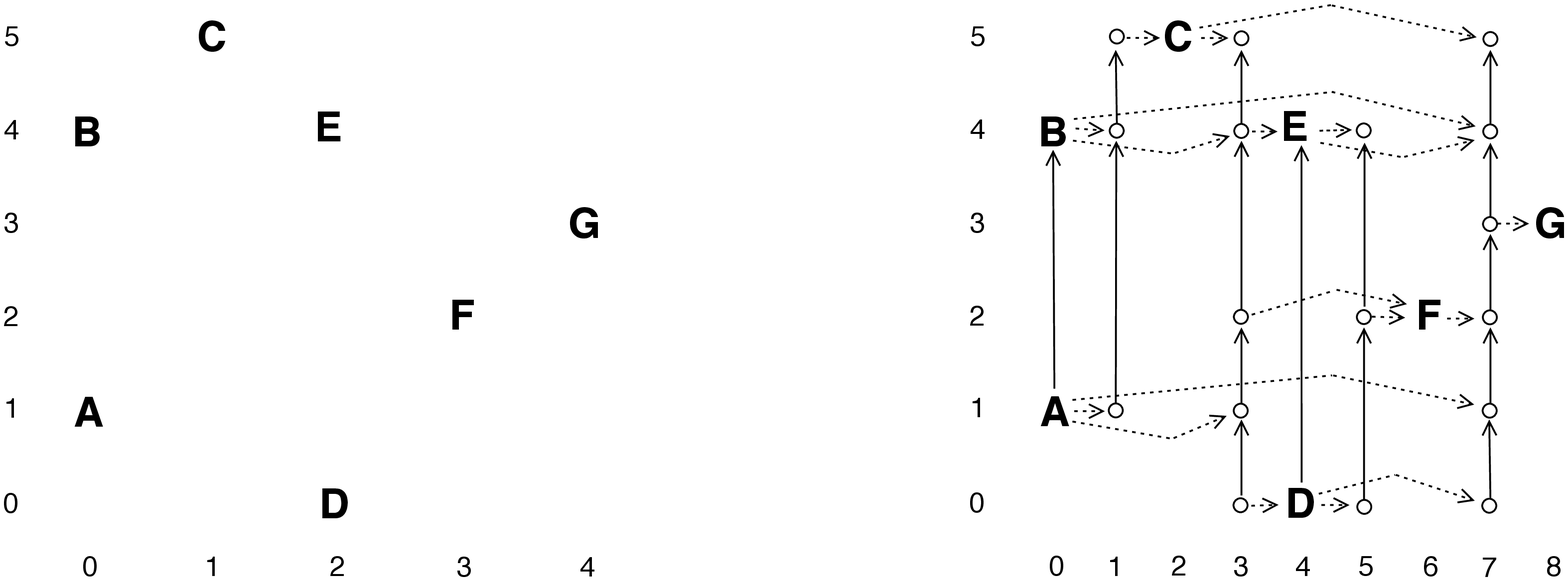}}}

\vspace*{0.3in} 
\hrulefill
\vspace*{-0.45in}

\caption{2-dimensional points in arbitrary position and their rendezvous lines}
\label{fig:Rendez2DLines}
\end{figure}

This observation was used in~\cite{QRendez} to create an order-preserving embedding of a set of $n$ points in $d$-dimensional space into a dag of $\Theta(n \log^{d-1} n)$ vertices and edges.
Modifying the construction slightly, let $a(i,h)$ be the index of the ancestor of $i$ at height $h$ in 1-dimensional rendezvous graphs and let $r(i,j)$ be the index of the rendezvous node for $i \leq j$ in 1-d rendezvous graphs (these do not depend on the size of the linear ordering as long as it is large enough for them to be defined).
For $d$-dimensional points $x=(x_1,\ldots,x_d)$, $y=(y_1,\ldots,y_d)$, if $x \prec y$ then a natural place for them to rendezvous is the $d$-dimensional vertex $(r(x_1,y_1), \ldots, r(x_d,y_d))$.
This was used for grids, but 
here they rendezvous at a line in the $d$th dimension, with its other coordinates being $r(x_1,y_1), \ldots r(x_{d-1},y_{d-1})$ (see Figure~\ref{fig:Rendez2DLines}).
$x$ is mapped to point $x_d$ on this line.

To determine $\eopt$ we need only find the maximum over all rendezvous lines of the maximum \meanerr{} on the line.
The lines are independent of each other and we do not use bounded envelopes from one for another.
Each point is mapped to $\lceil \lg n_1 \rceil \times \cdots \lceil \lg n_{d-1} \rceil$ lines.
The points are first sorted into lexical order $a(x_1,0)a(x_2,0) \cdots a(x_{d-1},0)x_d$ and the max \meanerr{} on the lines is determined, then into order $a(x_1,0)\cdots a(x_{d-1},1)x_d$, etc., as shown in Algorithm~\ref{alg:arbitrary}.

\begin{algorithm}
\setlength{\Ainindent}{0.4in}

\Ainnum{0}{1}$\mathsf{ \elow = 0 }$\\
\Ainnum{0}{2}$\mathsf{ for~ h_1 = 0 ~to~ \lceil \lg n_1 \rceil}$\\
\Ainnum{1}{3}$\mathsf{   for~ h_2 ~to~ \lceil \lg n_2 \rceil}$\\
\Ainnum{2}{4}$\mathsf{   \cdots }$\\
\Ainnum{2}{5}$\mathsf{     for~ h_{d-1} = 0 ~to~ \lceil \lg n_{d-1} \rceil }$\\
\Ainnum{3}{6}$\mathsf{      sort~ points~ into~ lexical~ order~ a(x_1,h_1)a(x_2,h_2) \cdots a(x_{d-1},h_{d-1})x_d}$\\
\Ainnum{4}{7}$\mathsf{      \{in~ case~ of~ ties,~ break~ them~ by~ sorting~ in~ x_1x_2\ldots x_d~ order\}}$\\
\Ainnum{3}{8}$\mathsf{      \elow = max\{\elow, max~ mean\_err~ on~ lines~ parallel~ to~ dimension~ d\}}$\\
\Ainnum{0}{9}$\mathsf{ construct~ isotonic~ regression~ using~ \elow,~ following~ a~ similar~ iterated~ sorting~ approach}$ 

\vspace*{0.3in} 
\hrulefill
\vspace*{-0.4in}

\caption{$L_\infty$ Isotonic regression for points in arbitrary positions in $d$-dimensional space} \label{alg:arbitrary}

\end{algorithm}

In Algorithm~\ref{alg:arbitrary} the structure in Figure~\ref{fig:Rendez2DLines} is never explicitly constructed.
Instead, sorting puts the vertices in a line into consecutive locations, and iterating through the sorting orders creates all of the rendezvous lines.
Thus the total space is only $\Theta(n)$ instead of the $\Theta(n \log^{d-1} n)$ needed in~\cite{QRendez} to create a multidimensional structure.
Each sort can be done in $\Theta(n)$ time by using radix sort (the implied constants depend on $d$),
and the maximum \meanerr{} on a line can be found in time linear in the number of points on the line (Algorithm~\ref{alg:linear}),
so the total time is $\Theta(n \log^{d-1} n)$.

There are some aspects in this process that need to be explicated.
While the maximum \meanerr{} on a line can be determined in time linear in the length of the line, $n_1$ may be superlinear in this length.
However, if a given line has $k$ points their actual first coordinate is ignored and they are treated as if the coordinates are $0,\ldots,k-1$.
Thus the time to find maximum \meanerr{} is $\Theta(k)$.
Further, on each line one only uses downward bounded envelopes from the small points rendezvousing here, and only upward bounded envelopes from the large points.
If $x=(x_1,\ldots,x_d)$ rendezvouses at line $(z_2\ldots,z_d)$, it is a small point iff $x_i \leq z_i$ for $1 \leq i \leq d-1$, and is large iff $z_i \leq x_i$ for $1 \leq i \leq d-1$.
If neither of these conditions hold then it is ignored.
All of the bookkeeping necessary to determine a point's rendezvous coordinates in each iteration, and whether it is small or large, can be done in $\Theta(1)$ time, where the implied constants depend on $d$.

This concludes the proof of Theorem~\ref{thm:points} b).

\section{Trees}  \label{sec:trees}

Recent applications of isotonic regression on trees include taxonomies and analyzing web and GIS data~\cite{Agarwaletal10,Chakrabartietal07,deKampetalClassificationTree,PuneraGhosh08}.
Here the isotonic ordering is towards the root.
For trees the previously fastest isotonic regression algorithms take $\Theta(n \log n)$ time for the $L_1$~\cite{QPartition}, $L_2$~\cite{PardalosXue99}, and $L_\infty$ metrics~\cite{QLinfty}.
We will show:

\begin{theorem}   \label{thm:tree}
Given weighted data \data\ on a rooted tree of $n$ vertices, an $L_\infty$ isotonic regression of the data can be determined in $\Theta(n)$ time.
\end{theorem}

The algorithm is based on a hierarchical decomposition.
First, given a rooted tree $T=(V,E)$ one can convert it to a rooted binary tree $T^\prime$ by replacing a vertex $v$ with $k>2$ children by a binary subtree with $k-1$ vertices.
The parent of the root of this subtree is $v$'s parent, and the $k$ external leaves are links to $v$'s children.
The number of vertices in $T^\prime$ is less than twice the number in $T$.
Data \data\ on $T$ it is extended to data on $T^\prime$ by assigning $y(v), w(v)$ to all vertices in the binary tree representing $v$.
An optimal isotonic regression of this data on $T^\prime$ yields an optimal isotonic regression on $T$ by assigning to $v \in V$ the value of the regression at the root of the subtree representing $v$ in $T^\prime$.
Because all of the above transformations can be done in linear time, from now on we assume the tree is binary.

The rendezvous graph is created incrementally, with each level being constructed from the one below.
Each rendezvous node $r$ covers a subtree of $S_r$ of $T$, where at most 2 of the leaves of $S_r$ aren't also leaves of $T$.
No two nodes at one level cover the same vertex in $T$, and thus 
each level partitions $T$ into subtrees and the nodes at that level form a binary tree.
This is similar to the partitioning for a linear order, where the partitioning was into subintervals.
The tree edges within a level are not used, but serve as a guide to construct the level above.

\begin{figure}
\centerline{\resizebox{5in}{!}{\includegraphics{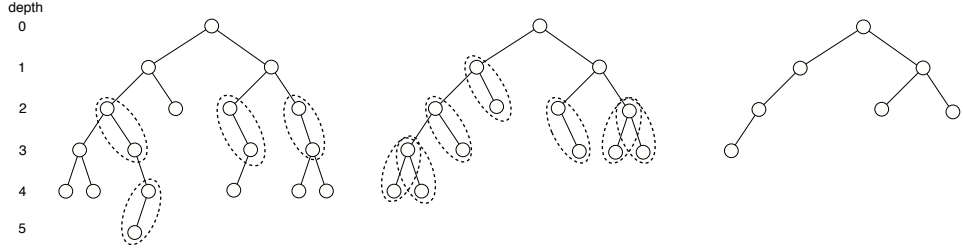}}}

\vspace*{0.3in} 
\hrulefill
\vspace*{-0.4in}

\caption{The two steps in going from one level to the next in rendezvous graphs for trees}
\label{fig:RendezTree}
\end{figure}

Given a tree $T_h$ at level $h$ of the rendezvous graph, to create the tree $T_{h+1}$ at level $h+1$,
\begin{enumerate}
\item \vspace*{-0.08in} First make a copy of $T_h$.
\item \vspace*{-0.08in} For every vertex $v$ of this copy, if $v$ has a single child then if $v$ is at an odd depth it merges with its parent and otherwise merges with its child.
\item \vspace*{-0.08in} In the resulting tree, every leaf merges with its parent, finishing the construction of $T_{h+1}$.
\end{enumerate}
\vspace*{-0.08in}See Figure~\ref{fig:RendezTree}.
This 2-step process is repeated level by level until only a single node remains.
Note that the rules insure that the tree remains binary at each level.

When nodes merge they are rendezvousing.
With $r$ is stored the downward envelop of all of the vertices in $S_r$ (the envelope could be viewed as being associated with the root of $S_r$), and for each of the $\leq 2$ leaf nodes of $S_r$ that aren't leaf nodes of $T$
it has the upward envelope of the vertices in $S_r$ on the path to the leaf.
There are $\leq \lfloor m/2 \rfloor$ nodes with 2 children, and all other nodes must merge with some node, hence the next level has $\leq 3m/4$ nodes (better bounds are possible), and thus the total size of the rendezvous graph is linear in the size of the original tree.
To determine the number of pairwise feasibility tests needed at each level, downward envelopes are merged at most twice, while upward envelopes are only merged once, so $k=2$ and $c=3/4$ in Observation 1,
so using 3 tests to reduce the number of segment endpoint errors suffices.

This proves Theorem~\ref{thm:tree}.

\section{Final Remarks}  \label{sec:final}

Isotonic regression is a fundamental nonparametric regression, making only a weak shape-constrained assumption.
It can be applied to linear and multidimensional orders without artificially requiring a metric on natural orderings such as S $<$ M $<$ L $<$ XL, and can be applied to more general ordering such as classification trees.
This makes it of increasing interest in machine learning and data mining~\cite{CaruanaNicul06,Chakrabartietal07,FeeldersRelabel2010,Gamarnik,Isotron,IntervalRank,PuneraGhosh08,RadeetalRelabel,VelikovaDaniels}.
The algorithms herein improve upon previous results by $\Omega(\log n)$ and are optimal for grids and trees.
This improvement also occurs for $d$-dimensional points in general position, an area where 
previous algorithms were criticized as  being too slow, forcing researchers to use inferior substitutes such as approximations or additive models~\cite{BurdakovetalGeneralPAV,Gamarnik,LussetalNIPS2010,SaarelaArjas,SalantiUlm,Sysoevetal2011}.

The algorithms use a mix of a nonconstructive feasibility test, rendezvous graphs, and bounded error envelopes.
The test and bounded envelopes are new to this paper.
The nonconstructive pairwise test allows one to move up the rendezvous graph, rather than continually returning to the base graph for the constructive test used previously.
Bounded error envelopes are important since standard error envelopes require $\Theta(n \log n)$ time and space just to build them~\cite{ChenWangPiecewise2013,GuhaShimLinftyHistogram,QRendez}.

Rendezvous graphs for isotonic regression on multidimensional points in arbitrary position were introduced in~\cite{QRendez} (a preliminary version was posted in 2008).
Two variants were introduced: one had a strong property that, given a set of $d$-dimensional points $P$, the rendezvous dag $R=(V,E)$ had $P \subset V$, and for any $p, q \in P$, $p \prec q$ iff there is a rendezvous node $r \in V\setminus P$ such that $(p,r)$, $(r,q)$ $\in E$.
If $p \not \prec q$ then there is no path in $R$ from $p$ to $q$, and thus the transitive closure of the domination ordering is represented by paths of length 2.
This is also known as a Steiner 2-transitive-closure spanner, and its size, $\Theta(n \log^d n)$, is optimal~\cite{SofyaetalSpanner}.
The second variant corresponds to the form used here, reducing the size to $\Theta(n \log^{d-1} n)$, though the transitive closure can involve paths of length $\Theta(n)$ (this can be reduced to $\Theta(\log n)$ by replacing lines with binary trees).
This is the smallest known dag where its vertices contain the original points and there is a path in the dag from point $p$ to point $q$ iff $p \prec q$.
For $d > 2$ the optimal size of such a dag is an open question.
Optimality for $d=2$ appears in~\cite{QRendez}.

Finally, the iterative sorting approach in Algorithm~\ref{alg:arbitrary}, which uses only $\Theta(n)$ space to provide the functionality of a multidimensional rendezvous dag with $\Theta(n \log^{d-1} n)$ edges, gives simple solutions to other multidimensional problems such the domination and empirical cumulative distribution function (ECDF) problems in~\cite{BentleyMultiDim}.
It gives the same time bounds but without needing modified algorithms for all lower dimensions.
The approach can also produce the transitive closure in $\Theta(n \log^{d-1} n + K)$ time and $\Theta(n+K)$ space, where $K$ is the number of edges in the transitive closure.
Further, all previous algorithms for $L_p$ isotonic regression on $d$-dimensional points, for any $1 \leq p \leq \infty$, required that the dag be given explicitly, and hence took $\Omega(n \log^{d-1} n)$ space.
The only exception is that one can do unweighted $L_\infty$ isotonic regression in $\Theta(n)$ space since it only utilizes simple operations that can be accumulated pairwise, but this approach takes $\Theta(n^2)$ time.

\end{document}